\documentclass[]{aa} 
\usepackage[varg]{txfonts}

\usepackage{natbib}
\usepackage{ulem}
\usepackage[breaklinks]{hyperref}    
\bibpunct{(}{)}{;}{a}{}{,} 

\makeatletter
\renewcommand*\aa@pageof{, page \thepage{} of \pageref*{LastPage}}
\makeatother

\defcitealias{Gough1966}{GT}

\begin{document}

\title{Characterization of magneto-convection in sunspots}
\subtitle{The Gough-Tayler stability criterion in MURaM sunspot simulations
    \thanks{videos associated with Figs.~\ref{fig:GT} \& \ref{fig:schwarz}
        available at \href{http://no.such.adress/exists/animation_2}{http://www.aanda.org}}}
\author{
    M. Schmassmann
    \inst{1}
       \and
       M. Rempel
       \inst{2}
       \and
       N. Bello {Gonz\'alez}
       \inst{1}
       \and
       R. Schlichenmaier
       \inst{1}
       \and
       J. Jur\v c\'ak
       \inst{3}
   }
\institute{Leibniz-Institut f\"ur Sonnenphysik (KIS), 
    Sch\"oneckstr. 6, D-79104 Freiburg i.Br., Germany\\
       \email{[schmassmann;nbello;schliche]@leibniz-kis.de}
       \and
  High Altitude Observatory, NCAR, P.O. Box 3000, Boulder, CO 80307, USA\\
        \email{rempel@ucar.edu}
        \and
        Astronomical Institute, Czech Academy of Sciences, Fri\v cova 298, 25165 Ond\v rejov, Czech Republic\\
        \email{jan.jurcak@asu.cas.cz}
  }
\date{Received: June 21, 2021; received with minor changes August 30, 2021; accepted September 13, 2021; TODAY: \today}

\abstract
  {
  Observations have shown that in stable sunspots, the umbral boundary is outlined by a critical value of the vertical magnetic field component. However, the nature of the distinct magnetoconvection regimes in the umbra and penumbra is still unclear.}
  {We analyse a sunspot simulation in an effort to understand the origin of the convective instabilities giving rise to the penumbral and umbral distinct regimes.}
  {We applied the criterion from Gough \& Tayler (1966, GT hereafter), accounting for the stabilising effect of the vertical magnetic field, to investigate the convective instabilities in a MURaM sunspot simulation.}
  {We find: 
  (1) a highly unstable shallow layer right beneath the surface extending all over the simulation box in which convection is triggered by radiative cooling in the photosphere;
  (2) a deep umbral core (beneath $-5$ Mm) stabilised against overturning convection that underlies a region with stable background values permeated by slender instabilities coupled to umbral dots;
  (3) filamentary instabilities below the penumbra nearly parallel to the surface and undulating instabilities coupled to the penumbra which originate in the deep layers. These deep-rooted instabilities result in the vigorous magneto-convection regime characteristic of the penumbra;
  (4) convective downdrafts in the granulation, penumbra, and umbra develop at about 2\,km\,s$^{-1}$, 1\,km\,s$^{-1}$, and 0.1\,km\,s$^{-1}$, respectively, indicating that the granular regime of convection is more vigorous than the penumbra convection regime, which, in turn, is more vigorous than the close-to-steady umbra;
  (5) the GT criterion outlines both the sunspot magnetopause and peripatopause,  highlighting  the  tripartite  nature  of  the  sub-photospheric layers of magnetohydrodynamic (MHD) sunspot models; and, finally,
  (6) the {Jur\v c\'ak} criterion is the photospheric counterpart of the GT criterion in deep layers.
  }
  {The GT criterion as a diagnostic tool reveals the tripartite nature of sunspot structure with distinct regimes of magneto-convection in the umbra, penumbra, and granulation operating in realistic MHD simulations.}

   
\maketitle
\section{Introduction}\label{sec:intro}


Sunspots are the most prominent manifestation of magnetic fields on the solar surface. They are characterised by a dark umbra surrounded by a brighter penumbra that exhibits elongated radial filaments. The fundamental physical process responsible for the clear distinction between umbra and penumbra is not fully understood and we focus on this process in this paper.

In the magnetohydrostatic sunspot models by \cite{1989A&A...222..264J} and \cite{1994A&A...290..295J}, a distinction between umbra and penumbra is made. They introduced the 'peripatopause'\footnote{The term is a derivative of the peripatetic Aristotle, who had the practice of walking to and from while teaching. In those models, the location of the peripatopause would wander until umbra and penumbra are in horizontal pressure equilibrium.} as the sheet that separates umbra and penumbra. In these `tripartite' (quiet sun, penumbra, umbra) models, the umbra is thermally isolated from the penumbra and the description of the convective heat transport in the two stratifications is distinct. The sheet between the magnetic penumbra and the non-magnetic quiet sun was introduced as the 'magnetopause' \citep{1970SoPh...13...85S,schmidt+wegmann1983}. These models were tuned to reproduce the observed umbral and penumbral brightness as well as the radial dependence of the photospheric magnetic field. While the heat transport in these models is restricted to the mixing length theory, they already assume that two distinct regimes of convection operate in the umbra and penumbra. In the tripartite model by \cite{1994A&A...290..295J}, the inclined heat-transmissive magnetopause is invoked to cause the surplus brightness of the penumbra by interchange convection of buoyant inclined magnetic flux tubes. While these models suggest that a critical inclination angle of the magnetic field lines at the magnetopause is needed to form a penumbra, they do not suggest a physical cause for the peripatopause.

One physical property that characterises the peripatopause in the photosphere was recently derived from observations:
the vertical component of the magnetic field, $B_\text{ver}$, 
at the umbral boundary of a stable sunspot is independent of spot size \citep{2011A&A...531A.118J, 2018A&A...611L...4J} and constant in time \citep{ 2018A&A...620A.104S}.
\citet{2018A&A...611L...4J} used 
Hinode/SP observations and found a constant
${|B_\text{ver}|=1867_{-16}^{+18}G}$ 
for a set of 144 umbral boundaries defined by $I_\text{c}=0.5\,I_\text{qs}$, 
that means where the continuum intensity, $I_{\rm c}$, is half that observed in the mean quiet sun, $I_{\rm qs}$.
\citet{2018A&A...620A.104S} used SDO/HMI observations to analyse a single stable spot during its disk passage for some nine days, concluding that 
$|B_\text{ver}|=1693\pm 15\,\text{G}$ at the umbral boundary is constant over time.
The difference of the $B_\text{ver}$ values is ascribed to differences in the spatial and spectral resolution, as found by
\citet{2017ApJ...851..111S}, who investigated the differences between HMI and SP data to find comparable differences for the magnetic field strength.
More detailed findings on the magnetic properties of the umbral-penumbral boundary can be found in
\citet{2011A&A...531A.118J},
\citet{2015A&A...580L...1J},
\citet{2017A&A...597A..60J},
\citet{Lindner2018},
\citet{2018A&A...620A.191B},
\citet{2020A&A...638A..28J}, and
\citet{2021A&A...649A.129G}.

\citet{Mullan2019} has pointed out that the invariant $B_\text{ver}$ at the umbral boundary of observed sunspots can be interpreted using the Gough-Tayler criterion \citep[][GT hereafter]{Gough1966}. The GT criterion expands the hydrostatic Schwarzschild stability criterion by including the stabilising effects of the magnetic field: the vertical component of the magnetic field stabilises against overturning convection, while the horizontal component of the magnetic field, $B_\text{hor}$, does not influence the stability but exclusively shapes the convective cells, typically into elongated structures  \citep{1961hhs..book.....C,2005MNRAS.360.1290T}. For more on magneto-convection in idealised setups, we refer to \cite{2005MNRAS.360.1290T} and the references therein, particularly \cite{2002MNRAS.337..293W}.
Therefore, even though the GT criterion is derived for simple field geometries, this criterion may prove useful to enhance our understanding on the origin and nature of the distinct magneto-convective regimes operating in sunspots.

In this paper, we follow up the ideas of \citet{Mullan2019} and we apply the Gough-Taylor criterion to the sunspot simulation of \cite{2015ApJ...814..125R}, which we  describe in Sect.~\ref{sec:simulation}. The Schwarzschild and GT criterion, described in Sect.~\ref{sec:gt}, were applied to our simulation. The results are presented and discussed in Sect.~\ref{sec:results} and our conclusions are summarised in Sect.~\ref{sec:discussions}.

\section{Sunspot simulations }\label{sec:simulation}

In a pioneering work by  \cite{2012ApJ...750...62R}, realistic magnetohydrodynamic (MHD) simulations of a sunspot with a penumbra were obtained by forcing the horizontal component of the magnetic field at the upper boundary of the simulation box to be stronger than in the force-free field case (parametrised by $\alpha>1$; the potential field corresponds to $\alpha=1$). One of the many successes of this simulation is a description of the magneto-convective nature of penumbral filaments, which is both self-consistent and in good agreement with observations. However, for simulations with $\alpha$ < 1.5, the penumbra is too narrow, and for $\alpha\ge1.5$, the magnetic field is unrealistically inclined at the umbral boundary compared to observations \citep{2020A&A...638A..28J}. Consequently, $B_\textrm{ver}$ is significantly too low at the umbral boundary \citep[1.36 kG for $\alpha$ = 2,][]{2020A&A...638A..28J}.

\begin{figure*}[!ht] 
    \centering
    \includegraphics[width=\textwidth]{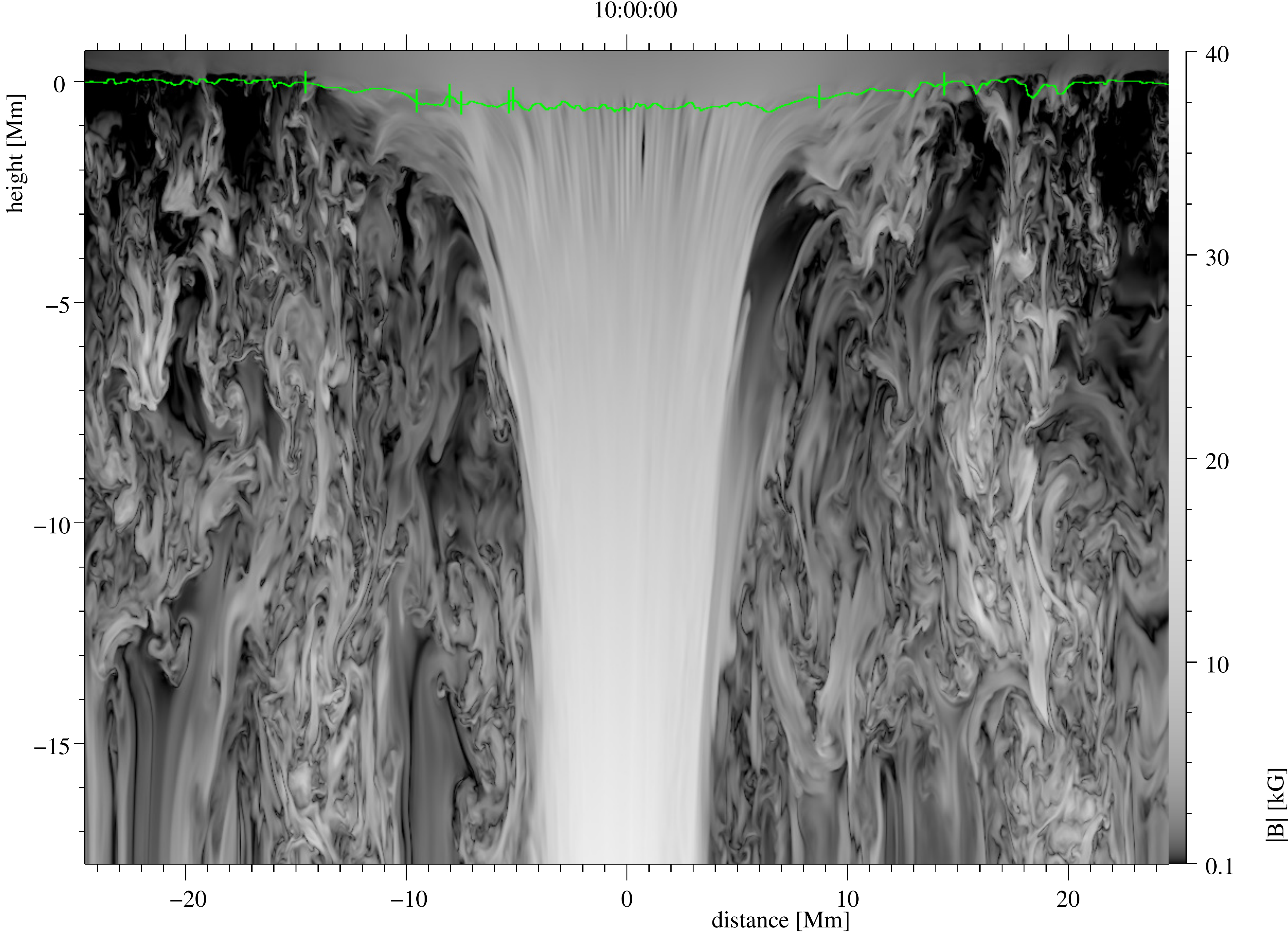}
    \caption{\label{fig:Babs} Cross-section of the total magnetic field strength of our simulation after 10\,h of solar time through the centre of the spot (same cut as in Fig.~\ref{fig:GT}). The sunspot magnetic field funnels out with its height embedded in the (granular) background magnetic field filling the box. 
    Vertical green lines mark the spot and umbral boundaries as seen in upper panel of Fig.~\ref{fig:GT}. The horizontal green line marks the $\tau\!=\!1$ surface.
    }%
\end{figure*}

For our analysis, we chose the MURaM sunspot simulation introduced in \cite{2015ApJ...814..125R}.
Our box has an extension of $98\times98\times18$\,Mm$^3$ with a horizontal (vertical) resolution of 48\,km (24\,km), but we only analyse  the centre $49\times49\times18$\,Mm$^3$. In this article, the vertical direction is denoted by $z$, with $z=0$ being the average height of the layer of optical depth unity of the quiet sun. Positive $z$ values are above optical depth unity in the photosphere.
The initial state is a small-scale dynamo simulation into which a self-similar magnetic field has been introduced. The total flux is $9\times10^{21}$\,Mx. The detailed parameters, as defined in Appendix A of \cite{2012ApJ...750...62R}, are: $B_0=20.25$\,kG, $R_0=4$\,Mm, and $z_0=13.38$\,Mm, resulting in a field strength on the spot axis at the top boundary of about 3\,kG.
The top boundary condition is a generalised non-potential field boundary condition, with $\alpha=2$ as described in Appendix B of \cite{2012ApJ...750...62R}.
The bottom boundary condition is MURaM standard, described in \cite{2014ApJ...789..132R} as $OSb$. 
The energy equation is simplified above the umbra of the sunspot. In regions with a density lower than $3\cdot 10^{-9}$ g cm$^{-3}$ the code omits numerical resistive heating, which is ill determined in this region due to the limitation of the Alfv{\'e}n velocity that is achieved through a reduction of the Lorentz force, as described in the Appendix of \cite{2009ApJ...691..640R}. In the following analysis, we mask out these regions.
For further details, including special treatment of the inner 15\,Mm in the first 5.56\,h, we refer to \cite{2014ApJ...789..132R}.

In Fig.~\ref{fig:Babs}, we present a cross-section of the total magnetic field strength of our simulation after 10\,h of solar time. In analytical models of monolithic sunspots, the magnetopause marks the outer boundary of the spot magnetic trunk. Naturally, in numerical spot models, this outer boundary is dynamic and corrugated, yet a clear drop in magnetic field strength is also present.

\section{Stability criteria}\label{sec:gt}

\subsection{Schwarzschild stability criterion against convection}\label{subsec:sc}

In the absence of magnetic fields, the Schwarzschild criterion for  stability against overturning convection is given by:
\begin{equation}
\nabla-\nabla_\mathrm{ad}\approx
\frac{1}{\Gamma_1}-\frac{\mathrm{d}_\text{z}\ln\rho}{\mathrm{d}_\text{z}\ln p}<0, \label{eq_S}
\end{equation}
with $p$ and $\rho$ denoting gas pressure and mass density, respectively, the first adiabatic exponent $\Gamma_1 := \left({\partial\ln p}/{\partial\ln\rho} \right)_\mathrm{ad}$, and
$\mathrm{d}_\text{z}=\mathrm{d}/\mathrm{dz}$ stands for a vertical derivative. As we treat every vertical column separately, the latter is a total derivative.
Using the term before the approximation is the original version from Schwarzschild, whereas the second version using $\Gamma_1$ and the logarithmic vertical derivatives of  $\rho$ with respect to $p$ correctly accounts for variations of mean molecular weight coming from variations in composition and ionisation. Although the later version is equivalent to what is sometimes called the Ledoux criterion \citep{1947ApJ...105..305L,2014A&A...569A..63G}, for simplicity and consistency with \cite{Gough1966}, we still refer to it as Schwarzschild criterion.
Here, we exclusively deal with the second version of the criterion. This inequality is a necessary and sufficient criterion for stability in the non-magnetic case \citep{Gough1966}.

\subsection{Gough and Tayler (GT) stability criterion}\label{subsec:gt}
All three-dimensional (3D) turbulent motions are hindered by magnetic fields \citep[e.g.][]{1974AN....295..275R}.
Because we are investigating the different types of mag\-neto-con\-vec\-tive processes in sunspot simulations, we need to include the effects of magnetic fields.
\cite{Mullan2019} related the observational invariant, $B_\text{ver}$, to the Gough-Tayler stability criterion \citepalias{Gough1966}, which in its simplest form is given by:
\begin{equation}
\frac{1}{\Gamma_1}-\frac{\mathrm{d}_\text{z}\ln\rho}{\mathrm{d}_\text{z}\ln p}-
\frac{{B_\text{ver}}^2}{{B_\text{ver}}^2+4\pi\Gamma_1{p}}<0.\label{eq_GTs}
\end{equation}
This and all other GT criteria are sufficient but not necessary criteria for stability.
Our notation differs from theirs, in that our fields are given in Gauss, and we identify their $\gamma$ as $\Gamma_1$, which is consistent with their definition of $\gamma$.
Allowing for $\Gamma_1$ to vary with height, while gravity is assumed constant, GT derive another criterion:
\begin{equation}
\frac{1}{\Gamma_1}-\frac{\mathrm{d}_\text{z}\ln\rho}{\mathrm{d}_\text{z}\ln p}-
\frac{{B_\text{ver}}^2}{{B_\text{ver}}^2+4\pi\Gamma_1{p}}
\left(1+\frac{\mathrm{d}_\text{z}\ln\Gamma_1}{\mathrm{d}_\text{z}\ln p}\right)<0.\label{eq_GTd}
\end{equation}
The section up to here is a summary of \citetalias{Gough1966}, with a different notation and identifying their $\gamma$ with $\Gamma_1$. Following the methods described in \citetalias{Gough1966} we also eliminated the derivative of $\Gamma_1$ by partial integration to obtain further criteria, but we do not describe it here, since it does not affect our results.

For our computation, we treat $\ln p$ as a function of $\ln u$ and $\ln \rho$, whereby $u$ is the internal energy per mass, and by defining adiabatic as
\mbox{$\mathrm{d}U+p\,\mathrm{d}V=0$}, or equivalently as 
\mbox{$\mathrm{d}\ln u-{p}/({\rho u}) \cdot \mathrm{d}\ln\rho=0$}, we get for $\Gamma_1$:
\begin{equation}
\Gamma_1=
\left(\frac{\partial\ln p}{\partial\ln\rho}\right)_\mathrm{ad}
=\left(\frac{\partial\ln p}{\partial\ln\rho}\right)_u+
\frac{p}{\rho u}
\left(\frac{\partial\ln p}{\partial\ln u}\right)_\rho,
\end{equation}
allowing the super-adiabaticity to be recast in the numerically well-conditioned form
\begin{equation}
\frac{1}{\Gamma_1}-\frac{\mathrm{d}_\text{z}\ln\rho}{\mathrm{d}_\text{z}\ln p}=
\left(\frac{\partial\ln p}{\partial\ln u}\right)_\rho
\frac{\mathrm{d}_\text{z}\ln u-\frac{p}{\rho u}\mathrm{d}_\text{z}\ln\rho}
{\Gamma_1\mathrm{d}_\text{z}\ln p}\, ,
\end{equation}
which also depends less on idiosyncrasies in the derivatives of the EOS tables.

\subsection{Applicability of GT criteria}

\citetalias{Gough1966} derived the criteria, Eqs.~\eqref{eq_GTs} and \eqref{eq_GTd}, in two simple setups: uniform inclined magnetic fields and horizontally varying vertical fields. 
Magnetic curvature forces are neglected and,
for instance, they do not consider effects such as the fluting instability. In addition, as the Schwarzschild criterion, the GT criteria assume that the stratification is in hydrostatic equilibrium.
In this strict interpretation, they only guarantee stability if the criteria are fulfilled throughout the full investigated volume. However, as shown in Sect. 4, the values in the inequality do provide a useful diagnostic to determine which parts of the simulation box are undergoing convection. This indicates (a posteriori) that dynamic effects and non-local magnetic forces are of minor importance when using the GT criteria to learn about overturning convection.

As mentioned above, the GT criterion for stability is sufficient, but not necessary. This means that stable regions (negative values) are identified unambiguously. But according to the derivation, regions with positive values are not proven to be unstable, they could still be stable. Yet, for practical purposes, we associate positive values above 0.01 with regions that are unstable. 
This can be justified  a posteriori by studying the temporal evolution of the  regions with values above 0.01. They are consistently found to be driven downwards.

\begin{figure*}[!ht] 
    \centering
    \includegraphics[width=\textwidth]{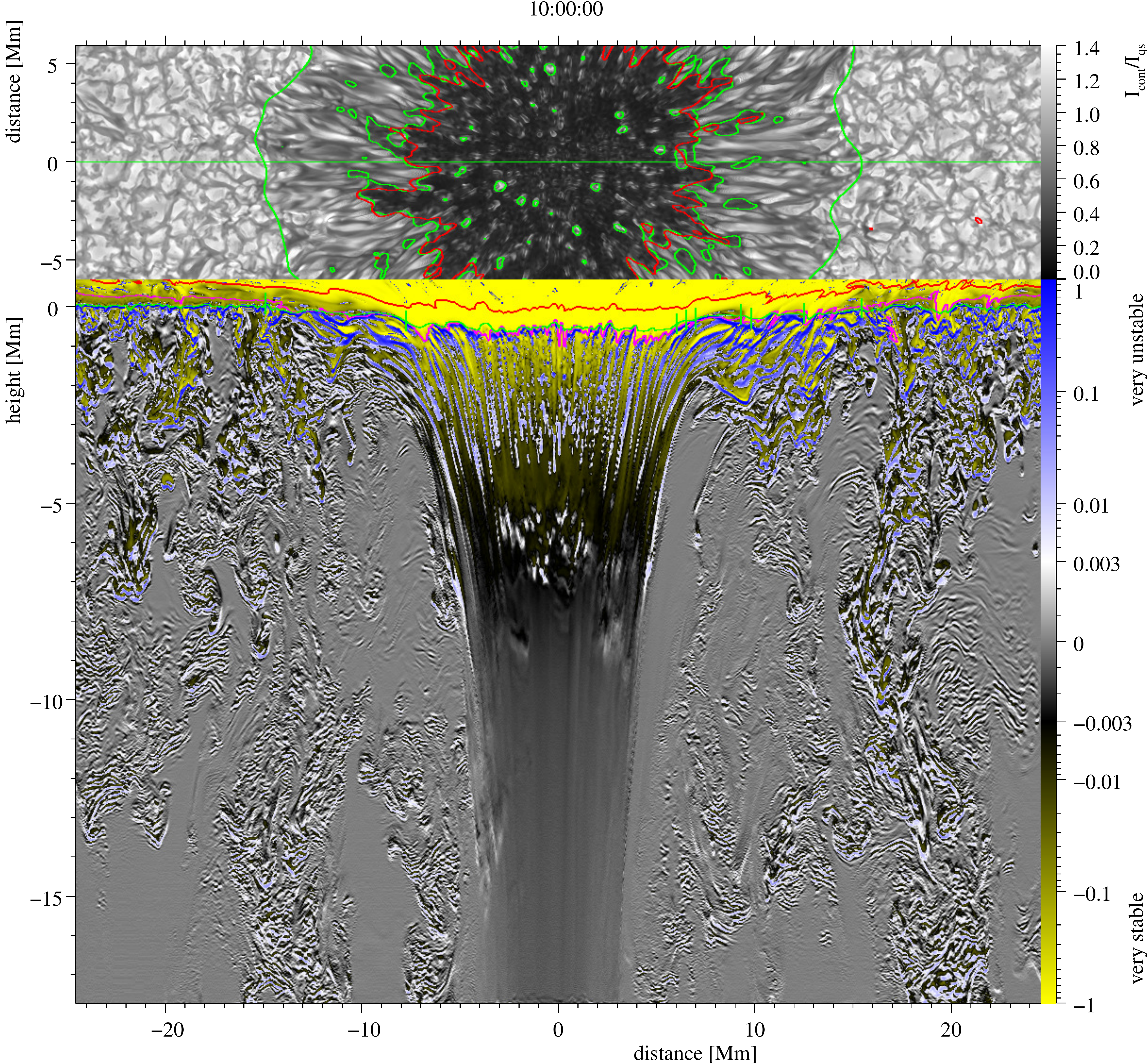}
    \caption {Sunspot showing partition into umbra, penumbra, and quiet sun as well as the magneto-convective stability of their sub-surface structure.
    Top: Bolometric intensity $I_c$.
    Bottom: Map of the GT stability shows the smaller of the LHSs of Eqs. \eqref{eq_GTs} and \eqref{eq_GTd} in a piece-wise logarithmic and linear colour scale from $-1$ to $+1$.
    The green lines are as in Fig.~\ref{fig:Babs}.
    The magenta contour marks the plasma $\beta = 8\pi p/ |B|^2 = 1$ surfaces.
    The red line marks the outer edge of the density $\rho<3\times10^{-9}\,\textrm{g}\,\textrm{cm}^{-3}$ region.
    The time evolution of this figure is available as animation \href{http://no.such.adress/exists/animation_2}{\bf online}.
    }
    \label{fig:GT}%
\end{figure*}

\begin{figure*}[!ht] 
    \centering
    \includegraphics[width=\textwidth]{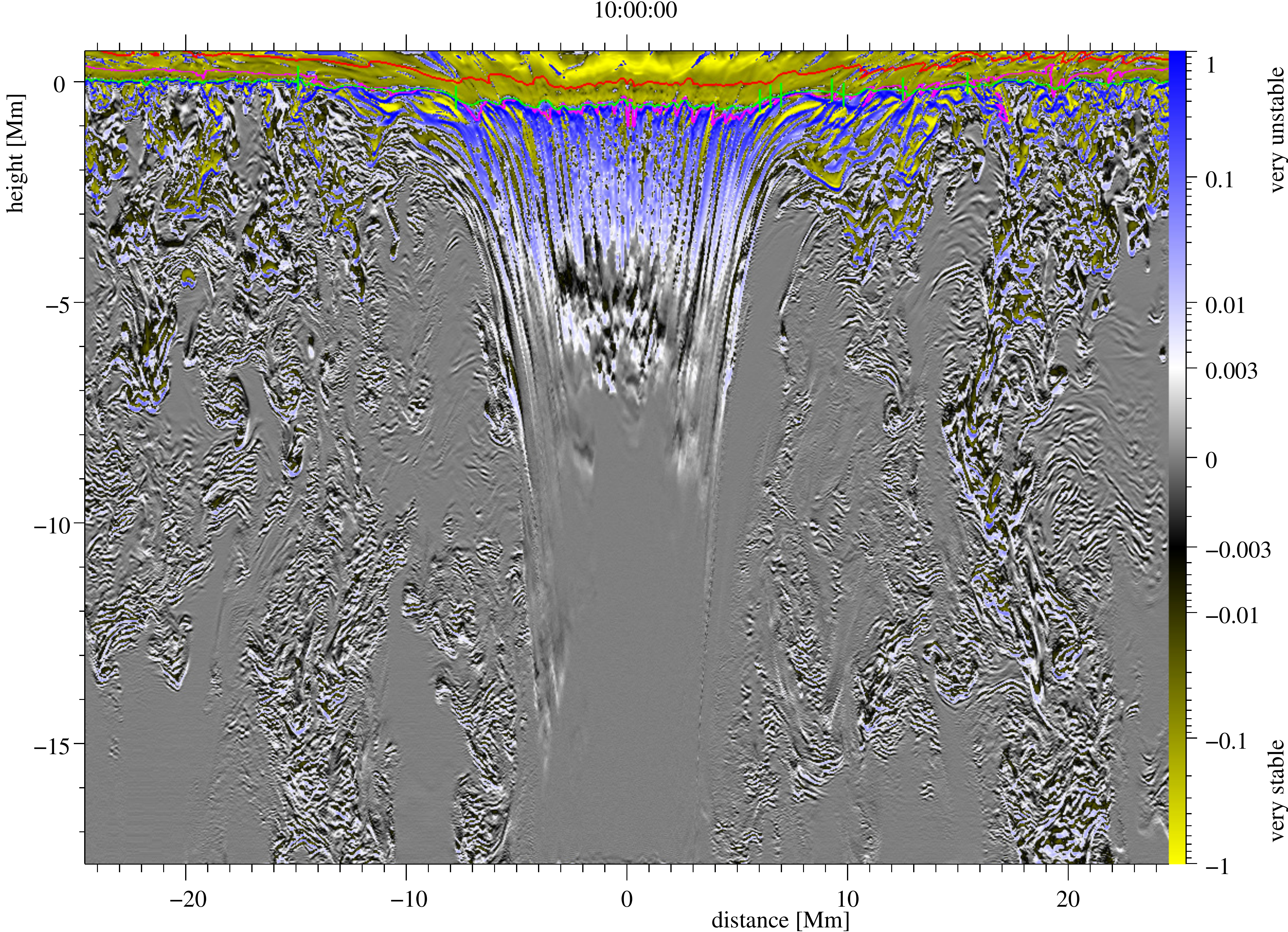}
    \caption {Calculation of the Schwarzschild criterion along a central cut of the MHD simulation run. Colour scale and lines identical to bottom panel of Fig.~\ref{fig:GT}. The time evolution of this figure is available as animation \href{http://no.such.adress/exists/animation_3}{\bf online}.
    }
    \label{fig:schwarz}%
\end{figure*}

\section{Results and discussion}\label{sec:results}

We determine the LHSs of the stability criteria (cf.~Sect.~\ref{subsec:gt}) in our simulation domain (cf.~Sect.~\ref{sec:simulation}) from 6\,h to 12\,h. The result for one snapshot at 10\,h and for a given cross section is presented in Fig.\,\ref{fig:GT}. The time evolution of this figure from 6\,h to 12\,h is available in an animation as \href{http://no.such.adress/exists/animation_2}{\bf online material}. We also calculate the Schwarzschild criterion (Eq.\,\ref{eq_S}) and present the co-temporal snapshot in Fig.~\ref{fig:schwarz}, for comparison.

The upper panel of Fig.\,\ref{fig:GT} shows the bolometric intensity, $I_\textrm{c}$, normalised to the mean quiet sun, with contours of $I_c=0.5\,I_\textrm{qs}$ in green and $B_\textrm{ver}=1867\,$G in red. To construct these contours, we smooth the maps with Gaussians that have a standard deviation of 144\,km. The outer green contour (defined by $I_c=0.9\,I_\textrm{qs}$ after smoothing with a Gaussian standard deviation of 768\,km) indicates the spot boundary and the central horizontal green line marks the position of the vertical cross-section of the lower panel (and of Fig.\ref{fig:Babs}).

We compute the LHS of the inequalities \eqref{eq_GTs} and \eqref{eq_GTd}. Both are sufficient criteria for stability. The minimum of both values for a given time is displayed in a vertical cut across the centre of the spot in the lower panel of Fig.\,\ref{fig:GT}. 

The colour code in the lower panel represents the degree of stability: (1) highly stable regions from $-$1 to $-$0.003 scale from yellow to black, (2) the transition from stable ($-0.003$) to unstable ($+0.003$) is in grey scale (from black to white), and (3) highly unstable regions from $+$0.003 to $+$1 scale from white to blue. 
The red contour outlines the density $\rho<3\times10^{-9}\,\textrm{g}\,\textrm{cm}^{-3}$ region in which the energy equation is not solved fully consistently, as described in Sec. \ref{sec:simulation}.

Although the Schwarzschild criterion is only valid in the absence of magnetic field, for illustration purposes, we compare it with the GT criterion. Figure~\ref{fig:schwarz} is equivalent to the lower panel of Fig.~\ref{fig:GT}, but computing the LHS of the Schwarz\-schild criterion, Eq.\,\eqref{eq_S}. In both figures, the photosphere above $\tau\!=\!1$ is found to be overall stable. While the Schwarzschild criterion yields unstable (blue and white) columns almost filling the volume below the umbra, the GT criterion marks this volume with a stable (yellow) background pierced by narrow unstable (blue and white) columns, 
as it takes into account the stabilising effect of the vertical magnetic field. Below the lower end of the GT unstable columns, $-4$ to $-5$\,Mm, there is GT stable and Schwarzschild marginally stable region. Below that, $-5$ to $-7$\,Mm, is a clumpy region, which is a leftover of a convective downdraft of the small-scale dynamo simulation used for the initialisation of this simulation. From this point to just above the bottom boundary, the structure is characterised by marginal Schwarzschild stability, hence, it is GT stable because of the stabilising effect of the magnetic field.

 The GT-stable spot magnetic trunk is embedded in a marginally stable (grey-coloured) background. As magnetic fields outside the spot trunk are small, Figs. \ref{fig:GT} and \ref{fig:schwarz} are almost identical there. Outside the spot and beneath the granular cells, black and white patches with shadings into yellow and blue outline the edges of granular convective cells, coincident with intergranular lanes at the surface. The animation (\href{http://no.such.adress/exists/animation_2}{\bf online material}) of the evolution at all times shows the dynamics of the GT criterion coupled to the continuous development of new instabilities. The effect of the vertical fields in intergranular lanes equally introduce a higher degree of stability compared to the Schwarzschild case. 
 
The highest GT instability values (in dark blue) are found right beneath the $\tau=1$ surface all along the simulation box.
These instabilities  develop, as it will be discussed later, due to plasma that has radiatively cooled and has increased density.

\begin{figure}[t]
    \centering
    \includegraphics[width=\linewidth]{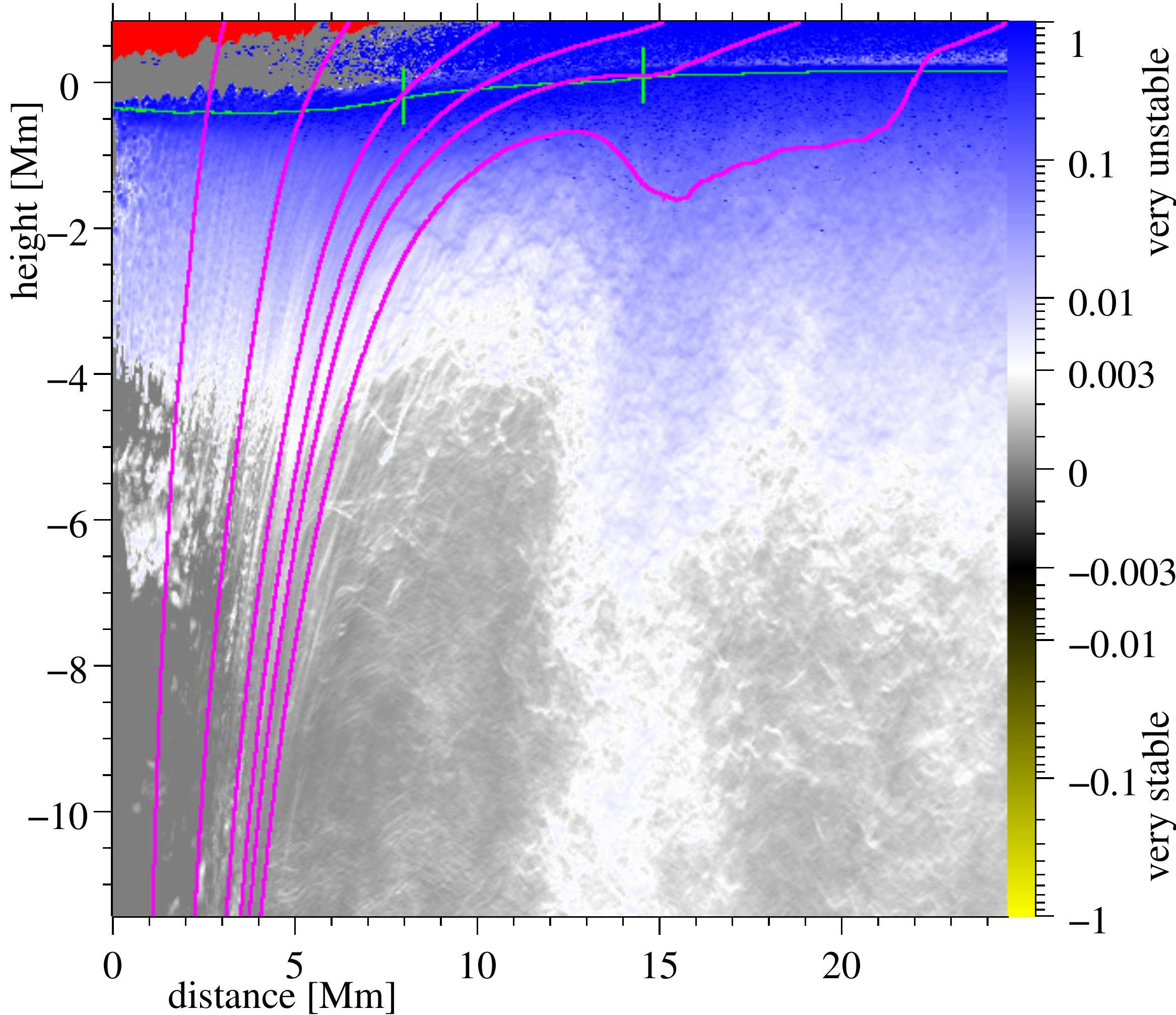}
    \caption{Same as lower panel of Fig.~\ref{fig:GT}, but here we plot the GT criterion as the azimuthal average over positive grid cell values around the spot. Cells with negative (stable) values are not considered. 
    The horizontal green line represents the azimuthal average of the $\tau=1$ surface. The vertical green lines mark the average position of the umbral and spot boundaries. In magenta we draw flux surfaces, that contain 7, 28, 51, 62, 68, and 74\% of the total flux respectively.
    \label{fig:pp} }%
\end{figure}

\subsection{GT criterion and nature of umbra and umbral dots}

In the thin unstable layer right beneath the umbral surface, radiative  cooling in the photosphere triggers slender-shaped instabilities that sink downwards. The unstable columns grow slowly downwards, reaching $-4\,$Mm after 5\,h. The penetration of these instabilities into deeper layers is slowed by the stable (yellow) core of the spot magnetic trunk. Such a stable stratification in the sub-photospheric umbra is expected to slow down any flow that penetrates into it, acting as an overshooting region for any radiatively cooled downflow. The overshooting convection depletes the energy reservoir below the umbra, which slowly changes the super-adiabaticity and thereby  erodes the stable region.
With increasing depth, the heat reservoir increases drastically, increasing the time-scale for this erosion.

This suggests that the convection in the umbra remains restricted to a surface layer operating above a core, stable against overturning convection. The unstable areas between $-5$ and $-7$\,Mm are leftovers from the quiet sun simulation, into which the magnetic trunk was set at the beginning of this sunspot simulation. Further below, plasma $\beta$ is low and the stabilising term in the GT instability gets lower, but remains sufficient to guarantee stability in the trunk. In the Schwarzschild criterion case, in which the stabilising effect of the magnetic field is not considered, these umbral instabilities are dominant below the umbra, and below $-8$\,Mm the stratification is marginally stable. 

These radiative-cooling driven instabilities populate the dark umbra with bright convective structures: umbral dots. They penetrate down to $-4$ Mm below the surface.
\cite{2006ApJ...641L..73S} reported on umbral dots permeating into the sub-surface by 1\,Mm depth at most. This shallow nature is possibly constrained by the smaller box size (1.6\,Mm depth) and shorter time runs (few hours) of their MHD umbra simulations.
The umbral surface layer of convection is constrained by an underlying stable stratification. This stable deep layer constitutes a heat reservoir as radiative heat transport is very inefficient. 

At the umbra-penumbra boundary, these thin instabilities below the surface penetrate deeper. We surmise that they could be the origin of peripheral umbral dots connected to bright penumbral grains, as seen in high-resolution observations \citep{2016A&A...596A...7S}.

\subsection{GT criterion and spot magnetopause}

The morphology of the  GT stability map (cf. lower panel of Fig.~\ref{fig:GT}) closely resembles  the spot magnetic trunk of Fig.~\ref{fig:Babs}. That means we find the GT criterion to be a good proxy to outline the overall sunspot magnetopause in the sub-photospheric layers.
In \cite{2020A&A...638A..28J}, we found that the highly corrugated magnetopause between the sunspot and the surrounding plasma coincides with the transition from super- to subequipartition field strength. Here, the magnetopause is marked by the stabilising effects of the vertical magnetic field strength.

\subsection{GT criterion and nature of sunspot penumbra}

Below the photospheric penumbra, the instabilities follow a filamentary structure. These filaments are connected to the magnetic trunk of the spot, where their lowest end reach deeper than those from the umbra. Deep below the spot, the penumbral instabilities are close to vertical. At the surface (green line) the umbral boundary marks a transition relative to the orientation of the surface: at the umbral boundary, they have the same inclination as the surface following the Wilson depression \citep{1774RSPT...64....1W}.
From there, they parallel the surface and become horizontal.
Also, below the penumbral surface, the filamentary instability structure is more horizontal than the azimuthally-averaged  shape of the field lines (see Fig.~\ref{fig:pp}) and undulate over time (\href{http://no.such.adress/exists/animation_2}{\bf online material}). This means that we see stable and unstable regions alternating in the GT maps down to a depth of $-3$\,Mm.

In some cases, where a penumbral filament is connected to a region of reduced intensity in the mid to outer penumbra, 
they develop a V-shaped instability (e.g. in Fig.~\ref{fig:GT} at height $-2$\,Mm, distance 10\,Mm). Due to radiative cooling, they are filled with low-entropy material and sink into deeper layers.
These V-shaped instabilities also develop continuously in the downdrafts of the surrounding granulation.

Some of these downdrafts drown to $-$8\,Mm depths. They seem to be related to their detachment from the spot itself into the convection zone at this stage of the sunspot simulation. At later times (after 80 hours; not shown here), when a moat flow has developed  \citep[see][]{2015ApJ...814..125R}, these very deep downdrafts only occur  outside the spot.

Penumbral, umbral, and granulation instability patches in the GT maps appear to evolve at different spatial and dynamical scales:
from an inspection of the evolution of the V-shaped downdraft fronts, we find that penumbral instabilities span over 3\,Mm (comparable to the length of penumbral filaments as seen in the bolometric intensity maps) and develop at a pace of $\sim$1\,kms$^{-1}$. Granular instabilities also develop a V-like shape, yet they are of a smaller size ($\sim$0.3\,Mm) than penumbral instabilities and drown faster at around 2\,kms$^{-1}$. The pattern propagation in penumbra und granulation are comparable to convective downflow velocities \citep[cf.][]{2012ApJ...750...62R}.

In the umbra, the subphotospheric stratification is stable initially.
There, slender umbral vertical downdrafts  with a width of 0.15\,Mm, at most, penetrate at 0.1\,kms$^{-1}$ down to some -4 Mm, where they saturate after 5h. The latter velocity is comparable to the convective rms velocity \citep[cf. Fig 7b in ][]{2012ApJ...750...62R} and may correspond to the downward propagation of the umbral cooling front.

The nearly one-to-one correspondence between the Schwarzschild and GT criteria maps in the penumbra and granulation indicate that the onset of downdrafts from the penumbra is more similar to that of the granulation than that of the umbra, in the sense that the reduced vertical magnetic field in the penumbra cannot stop convection from setting in. This is in agreement with the conclusions derived from the Jur\v c\'ak criterion \citep{2018A&A...611L...4J} as well as the theoretical studies by \cite{1961hhs..book.....C} on the effects of vertical fields in the inhibition of convection and the elongation of convective cells by the horizontal fields.

The penumbra can be thus understood as the transition regime between the umbra, in which the strong vertical magnetic field hinders convection, and the fully convective granulation. It is characterised by a convective nature filamented by the horizontal field of the spot.

\subsection{GT criterion and spot peripatopause}

\begin{figure}[t]
    \centering
    \includegraphics[width=\linewidth]{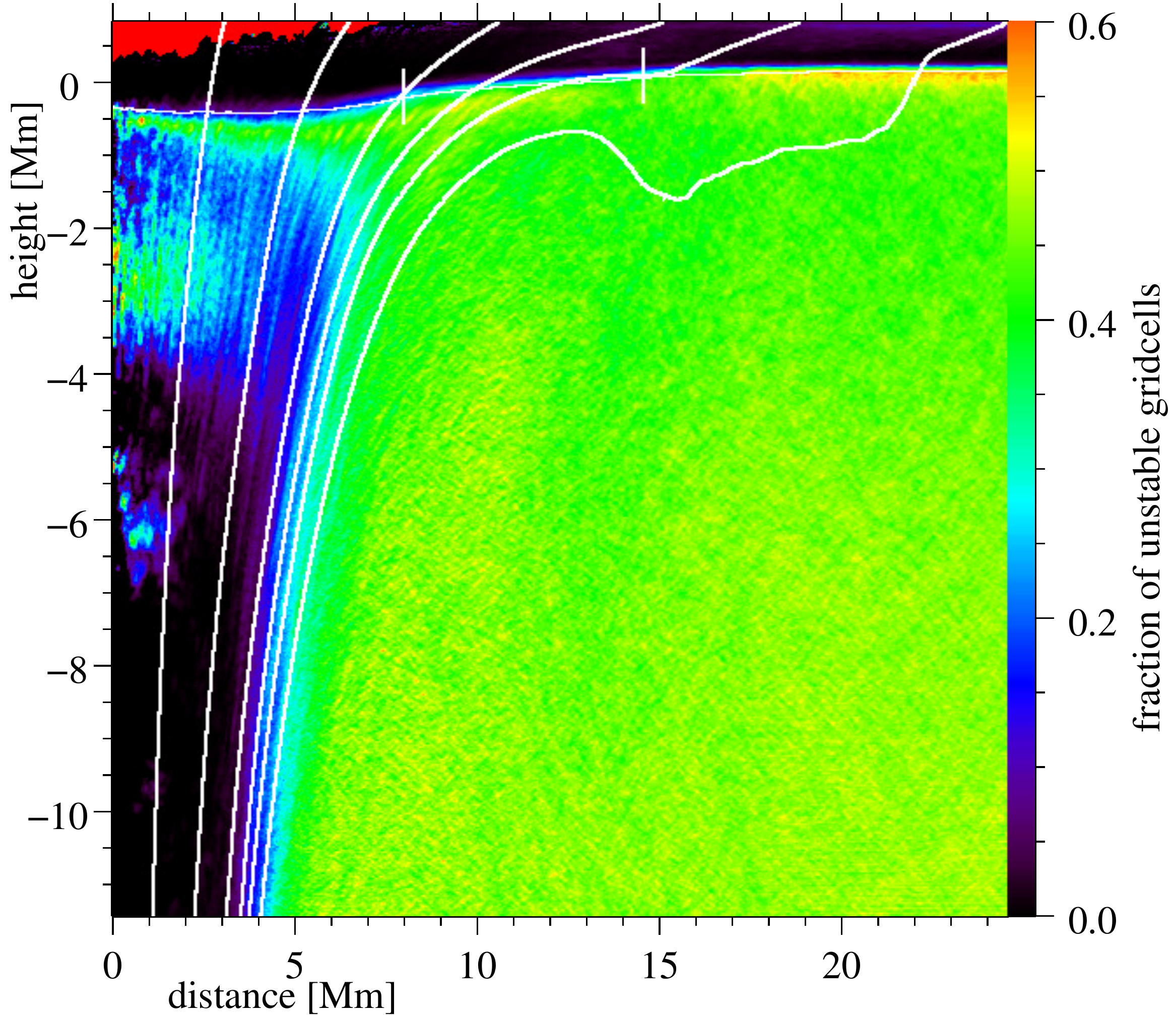}
    \caption{\label{fig:fill}
    Fraction of grid cells of a given height and radius that the GT criterion marks as positive, i.e. the filling factor of unstable gird cells. It is the same fraction of cells that are considered in the azimuthal average in Fig.~\ref{fig:pp}. Lines are as in Fig.~\ref{fig:pp}, but in white.
    }%
\end{figure}

We also find that the GT criterion makes a distinction between the umbra and penumbra convective regimes of the spot by outlining the presence of a peripatopause.
A way of highlighting the presence of the peripatopause is presented in Figs.~\ref{fig:pp} and \ref{fig:fill}. In Fig.~\ref{fig:pp}, we compute the azimuthal average of the GT criterion values and we display all grid cells that have a positive value (disregarding the negative ones) to assess a mean value of the instability at a certain radial distance and depth. In Fig.~\ref{fig:fill}, we show the corresponding filling fraction of positive (stable) grid cells. 

The magnetic flux surfaces shown in Figs.~\ref{fig:pp} and \ref{fig:fill} give the radius up to which we have to integrate to get the corresponding flux value. If the spot were spherically symmetric, these would correspond to field lines. 
The dip in the outermost flux surface, as well as the cells with a higher degree of instability at a radius of about 15\,Mm and below $-7$\,Mm, exhibits an enhanced drainage, which is an early indication of the moat cell, which starts to expand at this stage of the spot evolution. Inside that radius, the deeper layers are less unstable compared to the area outside.

In Fig.~\ref{fig:pp}, we see that the degree of instability is comparable for umbra, penumbra, and granulation; the closer to the surface, the stronger the instability. On the other hand, the filling fraction of unstable grid cells is lower below the umbra than in the surrounding granulation. For depths beneath some $-7$\,Mm, the transition from non-unstable cells to more than 40\% unstable cells occurs over about 2\,Mm and the flux surfaces connecting to the penumbral surface in the centre of that range.
As single penumbral filaments intrude far into an umbral background and negative grid cells are ignored, the jump in Fig.~\ref{fig:pp} occurs at the inner end of the inner most penumbral filament. Hence, the jump marks the innermost position of a corrugated peripatopause. The average location of discontinuity between stable umbra and unstable penumbra is at a larger radial distance. It is important to emphasise that, far from being a static sheet, the peripatopause shows a highly dynamical and permeable nature, as seen in horizontal cuts at various depths (not shown here).

This is a significative finding for understanding the outcome of a three-dimensional time-dependent simulation. The GT criterion, as a diagnostic tool, is able to sense areas of distinct regime of magneto-convection in the umbra and penumbra for layers deeper than $-7$\,Mm. Had the spot been initially placed where no downflow plumes are present, it might have been possible to detect this discontinuity at even higher layers in the Sun. In near-surface layers between $-4$\,Mm and the photosphere, both the umbra and penumbra are unstable. The near-surface convection in the umbra produces the umbral dots, while in the penumbra, the convection is more vigorous and extends into the deeper layers. Therefore, the GT criterion offers an explanation for the increased heat flux of the penumbra.

\subsection{GT criterion and the \texorpdfstring{Jur\v c\'ak}{Jurcak} criterion}

Figures~\ref{fig:pp} and \ref{fig:fill} also serve to establish the link between the GT criterion for stability and the Jur\v c\'ak criterion (cf.~Sect.~\ref{sec:intro}): The boundary between umbra and penumbra (peripatopause) is clearly visible for depth layers beneath $-7$\,Mm. An extrapolation of this discontinuity up to the photosphere along the predominant direction of the magnetic field lines marks the photospheric umbra-penumbra boundary, where the Jur\v c\'ak criterion is defined. 
This is also manifest in the lower panel of Fig.~\ref{fig:GT}: the innermost slender (white) instabilities connect the averaged peripatopause with the averaged photospheric umbral boundary. 

Both the GT and Jur\v c\'ak criteria highlight the role of the vertical component of the magnetic field as a stabilising parameter. While the GT criterion is able to define the peripatopause in the deep sub-photospheric layers, the Jur\v c\'ak criterion defines the boundary between umbra and penumbra in the photosphere. We can thus say that the Jur\v c\'ak criterion is the observable counterpart of the sunspot peripatopause.

\subsection{Comparison of GT criteria}

Comparing the results for the different GT criteria shows that the LHSs of Eqs.\,\eqref{eq_GTs} and \eqref{eq_GTd} 
almost always show the same sign. Differences only occur in the deep photosphere or directly beneath it. The lower end of the unstable region below the umbra is below a local minimum of $\Gamma_1$, hence, the $\Gamma_1$ gradient is increasing the stabilising effect of the magnetic field. Therefore, regions that are unstable according to Eq.\,\eqref{eq_GTd} occasionally protrude less deeply when computed with Eq.\,\eqref{eq_GTs}.
Only in rare cases does this difference extend to more than one grid cell.
A local maximum of $\Gamma_1$ lies at or above the upper end of these unstable regions, such that the stabilising effect of the $\Gamma_1$ derivative does not lower their upper boundary, but it decreases the value of the LHS of Eq.\,\eqref{eq_GTd} below that of Eq.\,\eqref{eq_GTs}, which is already negative. Hence, the region above the umbra is shown to be more stable by including the $\Gamma_1$ derivative. In summary, we find that the inclusion of the gradient of $\Gamma_1$ does not affect  the results in a qualitative way, thus, it is sufficient to analyse Eq.~\eqref{eq_GTs}.

\subsection{Summary}

Applying the Gough-Tayler criterion as a diagnostic tool in a realistic MHD simulation aiming to understand the sub-photospheric sunspot structure, we find that:

\begin{enumerate}[(a)]

\item All over the simulation box, convection is triggered by radiative cooling in the photosphere. Below the granulation surrounding the spot, these instabilities drown freely throughout the box without impediment.

\item The deep umbral core (beneath $-$5\,Mm) is sta\-bi\-li\-sed against overturning convection. The downwards penetration of instabilities is restrained as they enter this region of stable stratification. This hinders the efficient transport of energy, resulting in a dark cool umbra. 

\item Umbral dots appear to be the only type of magneto-convective cells driven by radiative cooling that develop in the umbra. They are connected to vertical slender columns of instability penetrating down to $-$\,4\,Mm. 

\item Meanwhile, the instabilities connected to the penumbra follow the field structure: near vertical where connected to the deep layers of the magnetic trunk, but close to parallel to the surface directly below the penumbra.

\item Umbra, penumbra, and granulation instabilities show distinct spatial and dynamical scales: while umbral instabilities are slender vertical columns developing at low speed (0.1\,km\,s$^{-1}$), penumbral horizontal instabilities develop a V-like shape of sizes comparable to penumbral filaments drowning at around 1\,km\,s$^{-1}$. Instabilities in granulation also develop a V-shape spanning over 0.3\,Mm and drowning faster ($\sim$2\,kms$^{-1}$) and deeper than those in the penumbra.

\item We find that the GT criterion outlines both the sunspot magnetopause (dynamical and permeable boundary of the magnetic spot trunk) and the peripatopause (dynamical and permeable transition between umbra and penumbra), highlighting the tripartite nature of the sub-photo\-spheric layers of MHD sunspot models in which three distinct magneto-convective regimes operate simultaneously.

\item While the GT criterion defines a boundary between umbra and penumbra in layers deeper than some $-7$\,Mm, the Jur\v c\'ak criterion defines that boundary in the photosphere. Both criteria discriminate based on the vertical component of the magnetic field. The Jur\v c\'ak criterion can be understood as an extrapolation of the GT peripatopause into the photosphere.
\end{enumerate}

\section{Conclusions}\label{sec:discussions}

In this work, we re-analyse a well-known numerical simulation of a sunspot in a $98\times98\times18\,\textrm{Mm}^3$ box with a horizontal (vertical) cell size of 48 (24) km. In this simulation, the presence of the penumbra is supported by implying non-potential more horizontal magnetic fields at the upper boundary of the box. Thus, the simulations exhibit umbra, penumbra, and granulation that are compatible with the intensity observations of sunspots.

In order to enhance our understanding of sunspot structure, we applied the Gough-Tayler criterion of stability as a diagnostic tool. Locally, the GT criterion takes into account the stabilising effects of the vertical component of the magnetic field. We gain new insights into our  understanding of the tripartite structure of sunspots and on fundamental implications for the energy transport in sunspots.
We confirm, a posteriori, that the GT criterion is a suitable tool when applied locally to identify the different regimes of magneto-convection, although the derivation only guarantees its validity as a global criterion.

All over the box, radiative cooling at the photosphere is the triggering mechanism behind the onset of the observed instabilities. Maps of GT stability values outline the magnetic trunk of the sunspot and the GT criterion detects distinct regimes of magnetoconvection in the umbra and penumbra. The deep sub-photospheric layers of the umbra are stable, such that umbral convection (umbral dots) is restricted to a 4\,Mm depth layer. The magneto-convection in the penumbra permeates into much deeper layers, but in contrast to the surrounding granulation, the instabilities occur in slender filaments. These filamented instability structures lie close to horizontal in the near-surface layers and become more and more vertical as they deepen in the sub-photosphere. In the surrounding granulation, plumes that form by radiative cooling are allowed to sink freely throughout the box.

In this context, the GT criterion highlights the tripartite nature of sunspots and the presence of a highly dynamic and permeable peripatopause and a magnetopause.
The photospheric Jur\v c\'ak boundary empirically observed between umbra and penumbra can be understood as the surface manifestation of the peripatopause separating two distinct regimes of sub-photospheric convection in the umbra and penumbra.
At the same time, the convective regime in the penumbra can be understood as a transition between those of the umbra and the quiet sun:
on the one hand, the convective cells in the umbra are slender and hindered by the strong vertical magnetic field and on the other hand, cells in the granulation are purely convective due to the close-to-absent magnetic field. The penumbral cells manifest a transition as they carry out a vigorous convection owing to the reduced vertical field in the penumbra, but turn filamented owing to the strong horizontal penumbral field.

\begin{acknowledgements}
We thank the reviewer for constructive comments which helped to improve our paper.
The simulation used was originally run with high-performance computing support from Yellowstone (\url{http://n2t.net/ark:/85065/d7wd3xhc}) provided by NCAR’s Computational and Information Systems Laboratory, sponsored by the National Science Foundation, under project NHAO0002 and from the NASA High-End Computing (HEC) Program through the NASA Advanced Supercomputing (NAS) Division at Ames Research Center under project s9025. This research has been partially supported through NASA contracts NNH09AK02I (SDO Science Center) and NNH12CF68C.
This material is based upon work supported by the National Center for Atmospheric Research, which is a major facility sponsored by the National Science Foundation under Cooperative Agreement No. 1852977. 
This research has made use of NASA's \href{https://ui.adsabs.harvard.edu/}{Astrophysics Data System.}
MS thanks HAO/NCAR’s visitor program for financing his stay at HAO.
We thank Damien Przybylski for providing us with the FreeEOS tables including $\Gamma_1$.
\end{acknowledgements}

\bibliographystyle{aa}
\bibliography{schmassmannAA_GT}

\end{document}